# Optimal Design of Hybrid AC/DC Microgrids

| H. LOTFI, A. KHODAEI | S. BAHRAMIRAD | M. BOLLEN |
|---|---|---|
| University of Denver | ComEd | Lulea Univ of Technology |
| USA | USA | Sweden |

**SUMMARY**

DC loads (such as computers, data centres, electric vehicle chargers, and LED lamps) and dc distributed energy resources (such as fuel cells, solar photovoltaics, and energy storages) are rapidly growing in electric power systems, so dc systems are being introduced as emerging practical investment solutions compared to traditional ac options. DC microgrids offer several advantages such as the elimination of need for synchronizing generators and easier supply of dc loads. Hybrid microgrids can benefit from the advantages of both ac and dc microgrid types. Moreover, there would be a huge reduction in the number of required power converters which would enhance the microgrid efficiency and reduce investment and operation costs. This paper introduces hybrid ac/dc microgrid as a viable solution compared to individual ac or dc microgrids and focuses on its planning. The objective of the hybrid microgrid planning is to minimize the microgrid total cost, including investment cost of distributed energy resources (DERs) and converters, operation cost of DERs, the cost of energy exchange with the utility grid, and the cost of unserved energy during the planning horizon. The economic viability of the microgrid planning is investigated in this paper and it is shown how the optimal DER generation mix, the type of feeders, as well as the point of connection of DERs to feeders can be determined by updating the traditional ac microgrid planning model. Numerical simulations on a test microgrid exhibit the merits of the proposed model.

**KEYWORDS**

Microgrid, critical load, distributed energy resource, hybrid microgrid, planning.

Hossein.Lotfi@du.edu

# 1. INTRODUCTION

Microgrids have attracted significant attention throughout the world in recent years as viable solutions in providing reliable, less expensive, high quality, and green electric energy. Moreover, microgrids could potentially improve the energy efficiency and reduce network congestion due to the proximity of generation and loads, thus benefiting both customers and the system as a whole [1, 2]. The aforementioned advantages have stimulated a broad research in various aspects of microgrids. Most prior studies have focused on microgrid operation, control, and scheduling [2, 3], and research on microgrid planning is rather limited. Available studies on microgrid planning investigate individual ac or dc microgrids. The study in [4] provides a comprehensive review of ac and dc technologies in microgrids and investigates various parameters as well as advantages and disadvantages of each technology. Several advantages of dc technology over the ac are mentioned. On the other hand, it is stated that most microgrids still use ac technology due to the need for more research on standardization of dc systems in spite of more attention towards dc systems. It is also emphasized that hybrid microgrids are more practical than individual dc microgrids. In [5], a model for energy management and operational planning of microgrids with PV-based active generators is presented, which includes a deterministic energy management system for microgrids composed of PV units, energy storage system, and gas microturbines. The integration of the proposed deterministic energy management method for business customers in the microgrid is further discussed. The study in [6] investigates the optimal planning and design of a hybrid microgrid which aims at minimizing the total planning cost by considering environmental issues such as carbon emissions. In order to assess the economical and operational performance of the proposed model, various DER mixes are compared.

Following the widespread use of dc loads, i.e., equipment with an internal DC bus, (such as data centres, LED lighting, electric vehicle chargers, and other electronic devices) and dc DERs, i.e., generation sources with an internal DC bus (such as solar PV and fuel cell), there have been more attention towards dc microgrids. There are several advantages associated with dc microgrids such as eliminating the need for synchronization of distributed generators, easier integration of dc DERs, and easier supply of dc loads by eliminating multiple ac-dc-ac converters. In reality, however, both ac and dc loads/DERs exist, and it is more economical to have both ac and dc buses/lines which reduce the conversion costs to a high extent. Microgrid planning is an important problem that requires the simultaneous consideration of short-term operation and long-term investment problems, as well as the inclusion of several sources of uncertainty, such as errors in load forecast, renewable generation, electricity market price, and microgrid islanding. The microgrid planning problem is modelled in [7] by adopting robust optimization, and a model for ac microgrid planning is proposed that uses Benders decomposition method to decompose the planning problem into an investment master problem and operation subproblem. Uncertainties in microgrid operation and planning are also studied in [8] by proposing a preprocessing approach in lieu of robust optimization, which leads to a significant reduction in computation time. The study in [3] investigates individual ac or dc microgrid planning with the objective of determining the optimal generation mix of DERs and the type of microgrid. Also, the threshold ratio of dc loads, which makes dc microgrid a more economical solution compared to ac microgrid is found.

This paper is an extension to [3] by adding the capability for the microgrid to have both ac and dc buses/feeders as a more practical solution. A number of feeders and various types of distributed generations (DGs) are considered in this study. The objective is to minimize the microgrid total planning cost. It also determines the optimal DER generation mix, the point of connection of each DER to feeders, and the type of each feeder, i.e., either ac or dc, according to the ratio of dc loads in feeders and the ratio of critical loads. The economic viability of the proposed microgrid planning is examined by comparing the microgrid planning cost and the cost of supplying all loads with the utility grid, i.e., without microgrid deployment.

The rest of the paper is organized as follows: Section 2 investigates the model outline and problem formulation of hybrid microgrid planning; Section 3 provides numerical simulations on a test microgrid; and Section 4 concludes the paper.

## 2. HYBRID MICROGRID PLANNING MODEL OUTLINE AND FORMULATION

A general structure of a hybrid ac/dc microgrid is represented in Figure 1. A hybrid microgrid has several ac/dc feeders, but only one ac and one dc feeder are shown here. The ac DERs (including wind turbine, microturbine, etc.) and ac loads can be connected to ac feeders via a transformer (to convert the ac voltage to the desired level) or to dc feeders via a rectifier/inverter. Although these units may use ac/dc/ac conversion and would be very suitable for a dc collection grid, in this work they are considered as ac DERs, which however, can be modelled and used as dc DERs in the proposed formulations without loss of generality. Similarly, dc DERs (including fuel cell, solar PV, etc.) and dc loads can be connected to dc feeders via dc-to-dc converters or to ac feeders via an inverter/rectifier. To minimize the total cost in this configuration, ac components should be connected to ac feeders and dc components to dc feeders. However, in order to increase the system reliability, the components could be connected to both feeders at the same time as shown in the figure; so, if there is any problem with one of them, the other one can still support the loads/DER. All dc (ac) feeders should be connected to a common dc (ac) bus. The dc bus is to be connected to the point of common coupling (PCC) via a bidirectional converter (it is bidirectional since the exchanged power with the utility grid could be either positive, when importing power, or negative, when exporting power). At the end, there is a need for a transformer to connect the microgrid to the utility grid in order to have the desired voltage level.

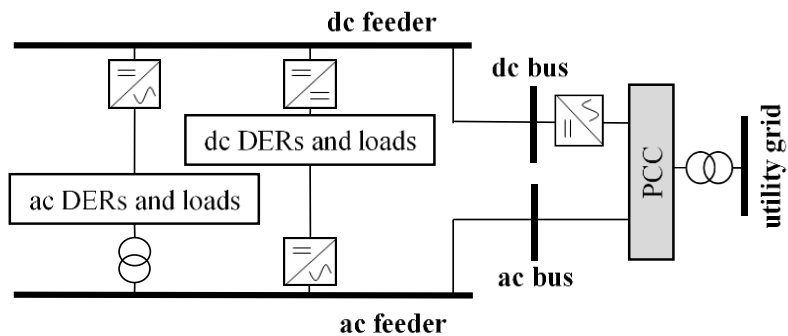

Figure 1. General structure of a hybrid ac/dc microgrid

Generally, one of the advantages of microgrid deployment is the use of renewable energy sources such as wind, solar, fuel cell, biomass, etc. Renewable DGs have been widely considered by utilities and prosumers due to their decreasing costs, governmental policies to reduce carbon emission, and rapid construction. Since renewable energy sources produce variable generation due to weather conditions, they should be accompanied by a distributed energy storage (DES) to store their energy whenever they generate power and supply the loads later during peak hours or microgrid islanding. The DES also improves the system economics by storing energy in low-price hours and selling it back to the utility grid in high-price hours under net metering contacts which increases the microgrid revenue. Moreover, there are critical loads in the microgrid that should be seamlessly supplied which necessitate a number of dispatchable DGs. As a result, both dispatchable and renewable DGs are considered in the microgrid planning. In the proposed model, it is assumed that a number of feeders exist that their type, i.e., ac or dc, and the connection of DERs to feeders should be determined by solving the microgrid planning problem.

The microgrid planning problem aims at minimizing the total planning cost which is comprised of the investment cost of all components (including DERs and converters), the operation cost of dispatchable DGs and energy exchange with the utility grid, as well as the cost of unserved energy, or the reliability cost. The investment cost is calculated annually while the operation and reliability costs are calculated hourly in the project lifetime. The calculation of the operation and reliability costs can be found in the literature [3]. The microgrid investment cost, however, is the term that is different from the existing literature when a hybrid microgrid is to be deployed. The investment cost can be represented as in (1):

$$IC = \sum_t \sum_k \sum_{i \in G} C_{1,ikt} x_{ik} + \sum_t \sum_k \sum_{i \in G_{ac}} C_{2,ikt} x_{ik} z_k + \sum_t \sum_k \sum_{i \in G_{dc}} C_{3,ikt} x_{ik}(1-z_k) + \sum_t \sum_k C_{4,kt} z_k + \sum_t \sum_k C_{5,kt}(1-z_k) \quad (1)$$

The investment cost includes five terms as discussed below:
$C_1$: The investment cost of DERs,
$C_2$: The cost of rectifiers required for connecting ac DERs to dc feeders,
$C_3$: The cost of inverters required for connecting dc DERs to ac feeders,
$C_4$: The cost of inverters for connecting ac loads to dc feeders, and
$C_5$: The cost of rectifiers for connecting dc loads to ac feeders.

Indices $t$, $i$, and $k$ denote year, DERs, and feeders, respectively. G denotes the set of DERs. $x_{ik}$ is the binary decision variables associated with the connection of DER $i$ to feeder $k$ when it is set to one. $z_k$ is another binary variable to determine the type of feeder $k$. The feeder is dc if $z_k$ is set to one, and ac if set to zero. Using the proposed investment cost, when ac DERs are selected to be installed, their associated investment cost will be added to the objective along with a rectifier cost if connected to a dc feeder. Similarly, when dc DERs are selected to be installed, their associated investment cost will be added to the objective along with an inverter cost if connected to an ac feeder. This is also true for ac and dc loads, where proper inverter/rectifier cost is added to the objective if they are connected to an opposite type feeder. All of these different cases will be obtained based on different combinations of $x$ and $z$ values as investment states.

The other constraint that needs to be revised in the microgrid planning problem is the load balance, as proposed in (2):

$$\sum_{i \in G_{ac}} (P_{it} + P_{M,kt})(\eta_{rec} z_k + (1-z_k)) + \sum_{i \in G_{dc}} P_{it}(z_k + \eta_{inv}(1-z_k)) = D_t \qquad \forall k, \forall t \quad (2)$$

The load balance constraint ensures that the sum of power from all DERs and the exchanged power with the utility grid in all hours is equal to the total load in the microgrid. $P_i$ and $P_M$ represent the hourly power generated by DERs and the exchanged power with the utility grid, respectively. $P_M$ can be either positive, when buying power from the grid, or negative, when selling the excess power to the grid. The efficiency of rectifiers and inverters, as shown by $\eta_{rec}$ and $\eta_{inv}$, respectively, is considered in modelling this constraint as there will be losses due to the application of converters. According to (2), the power generated by ac DERs located in dc feeders is reduced by the coefficient $\eta_{rec}$, and the power generated by dc DGs located in ac feeders is reduced by the coefficient $\eta_{inv}$. Again, the feeder type binary variable $z$ is used to impose proper terms of the constraints as the microgrid type is determined.

The microgrid planning problem is further subject to additional constraints, such as the charging and discharging constraints of DES, maximum power exchange with the utility grid, the technical constraints of dispatchable DGs, and the load adjustment and curtailment. These constraints, however, can be found in the literature [3] and used along with the proposed constraints to obtain the hybrid microgrid planning model.

## 3. NUMERICAL RESULTS

The proposed hybrid microgrid planning problem is applied to a test microgrid consisting of four dispatchable units (gas-fired), one wind unit, one solar PV unit, and one DES as represented in Table 1. The peak load is 8.5 MW. All load, solar, and wind generation, as well as the hourly market price are obtained from the Illinois Institute of Technology Campus microgrid in [3]. Three feeders are considered in this study and the planning horizon is assumed to be 20 years. The problem is formulated by mixed-integer programming (MIP) and solved by CPLEX 12.6. First, the ratio of dc loads and critical loads are considered to be 0.4 and 0.5, respectively. The ratios of dc (ac) loads to total dc (ac) loads in the microgrid in feeders 1, 2, and 3 are assumed to be 0.39 (0.28), 0.33 (0.33), and 0.28 (0.39), respectively.

Table 1 DGs Characteristics

| Unit No. | Type | Annualized Investment Cost ($/MW) | Allowable Installation Capacity (MW) | | Cost Coefficient ($/MWh) |
|---|---|---|---|---|---|
| 1,2 | Gas | 50,000 | Totally 5 | 2 | 85 |
| | | | | 1.5 | 95 |
| | | | | 1.5 | 105 |
| 3,4 | Gas | 70,000 | Totally 5 | 1 | 65 |
| | | | | 1 | 70 |
| | | | | 1 | 75 |
| 5 | Wind | 132,000 | 2 | | 0 |
| 6 | Solar PV | 133,000 | 2 | | 0 |

The hybrid microgrid planning solution would install all DGs with capacity of 0.049 MW for units 1 and 2, 2.376 MW for units 3 and 4, as well as 2 MW for wind and solar PV units. All feeders would be ac. The investment, operation, and total planning costs are $8,379,934, $2,737,252, and $11,117,190, respectively. Since the microgrid planning cost is less than the cost of supplying loads witout microgrid deployment ($14,548,920), the microgrid installation is economically viable. The sensitivity analysis with respect to the ratio of dc loads shows that by increasing this ratio up to 0.6, all feeders are ac. The increase in this ratio up to 0.8 would cause feeders 1 and 2 to be dc, and increasing to 0.9 and 1 will select all feeders to be dc. The microgrid costs are shown in Figure 2. By increasing this ratio to 0.8, the total dispatchable capacity and their hourly power generation increase, so the investment and operation costs increase as well. Having all loads as dc (associated with dc load ratio=1) causes the investment cost to decrease due to the elimination of the converters' costs. Moreover, the power loss reduction leads to less power from DGs, so the operation cost drops. Therefore, the planning cost would be an ascending curve first, and a descending one afterwards. By increasing the ratio of critical loads, the capacity of dispatchable DGs would increase, which causes the microgrid investment and planning costs to increase as well. Changing the electricity market prices from -10% to +20% also increases the total capacity of dispatchable units since the microgrid would sell more energy to the utility grid, which increases its revenue. As a result, the investment cost increases while the operation cost decreases by increasing the market price, as shown in Figure 3. The planning results show that solar and wind units are installed with their maximum capacity in all cases, conceivably due to their zero operation cost. Moreover, ac DERs are installed in ac feeders and dc DERs in dc feeders in order to avoid the conversion costs.

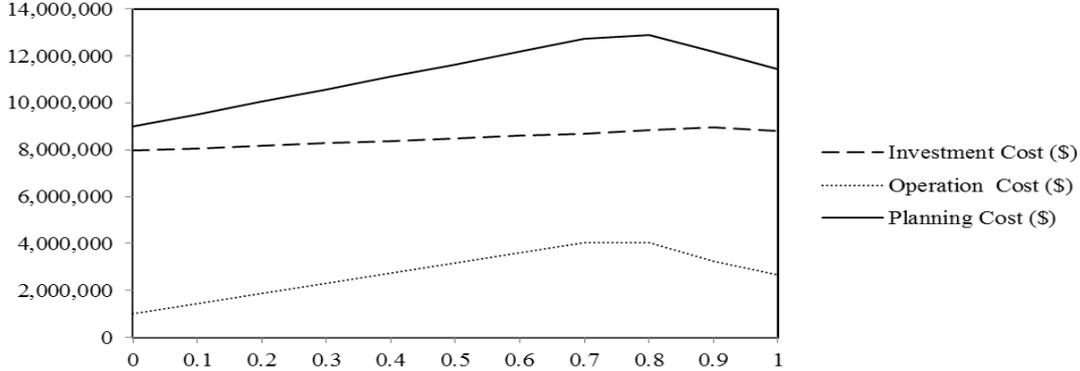

Figure 2. Microgrid costs with respect to ratio of dc loads to total load

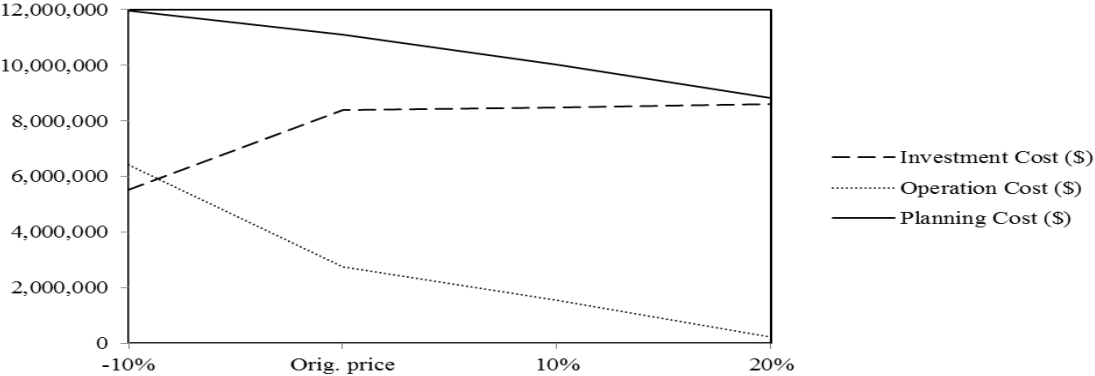

Figure 3. Microgrid costs with respect to market prices

## 4. CONCLUSION

In this paper, hybrid ac/dc microgrids' advantages, their role in improving system efficiency, and their various components were explained and a model for hybrid microgrid planning was presented with the objective of minimizing the total planning cost. The optimal size of DERs, their point of connection to feeders, and the feeders' type were determined as the output of the hybrid microgrid planning solution. The numerical simulations on a test microgrid demonstrated that the only factor in determining the type of feeders, i.e., either ac or dc, would be the ratio of dc loads. It was verified that increasing the ratio of critical loads would install a larger capacity of dispatchable units in order to seamlessly supply critical loads, which increases the investment, and hence, the planning costs. It was further shown that by increasing market prices, it would be more desirable for the microgrid to sell as much power as possible to the utility grid which requires installing more DG capacities, thus increasing the investment cost but decreasing the operation cost. One issue that needs more investigation is that many devices use a variable dc voltage and also often a number of different dc voltages internally. For example, a PV uses the dc voltage for the maximum power point tracking. When connecting directly to a dc grid, the voltage is no longer under control and the efficiency will be lower. Although this doesn't change the proposed methodology, but the conclusions may be different as even for dc devices like PV an additional dc/dc converter would be needed.